\documentclass[
aps,
prd,
groupedaddress,
nofootinbib, 
amsmath,
amsfonts,
preprintnumbers,
showpacs,
9pt,
english
]{revtex4-2}
\usepackage{amsmath}
\usepackage{amssymb}
\usepackage{babel}
\usepackage{wrapfig}
\usepackage{cancel}
\usepackage{dblfloatfix}
\usepackage{float}
\usepackage[none]{hyphenat}
\usepackage{placeins}
\usepackage{amsthm}
\usepackage{placeins}
\usepackage{caption}
\usepackage{booktabs}
\usepackage{array}
\usepackage{booktabs}
\usepackage{tabularx}
\usepackage{booktabs}
\usepackage{relsize,exscale}

\newcommand{\beq}{\begin{equation}} 
\newcommand{\eeq}{\end{equation}}
\newcommand{\bea}{\begin{eqnarray}}
\newcommand{\eea}{\end{eqnarray}}

\newcommand{\dd}{\text{d}}
\newcommand{\comm}[1]{}

\usepackage{array,multirow,graphicx}
\usepackage{dcolumn}
\usepackage{newlfont}
\usepackage{bm}
\usepackage[colorlinks,citecolor=blue,urlcolor=blue,linkcolor=blue]{hyperref}
\usepackage[figtopcap]{subfigure}
\usepackage{color}

\usepackage{scalerel}
\usepackage{tikz}
\usetikzlibrary{svg.path}
\definecolor{orcidlogocol}{HTML}{A6CE39}
\tikzset{
  orcidlogo/.pic={
    \fill[orcidlogocol] svg{M256,128c0,70.7-57.3,128-128,128C57.3,256,0,198.7,0,128C0,57.3,57.3,0,128,0C198.7,0,256,57.3,256,128z};
    \fill[white] svg{M86.3,186.2H70.9V79.1h15.4v48.4V186.2z}
                 svg{M108.9,79.1h41.6c39.6,0,57,28.3,57,53.6c0,27.5-21.5,53.6-56.8,53.6h-41.8V79.1z M124.3,172.4h24.5c34.9,0,42.9-26.5,42.9-39.7c0-21.5-13.7-39.7-43.7-39.7h-23.7V172.4z}
                 svg{M88.7,56.8c0,5.5-4.5,10.1-10.1,10.1c-5.6,0-10.1-4.6-10.1-10.1c0-5.6,4.5-10.1,10.1-10.1C84.2,46.7,88.7,51.3,88.7,56.8z};}}
\newcommand\orcid[1]{\href{https://orcid.org/#1}{\mbox{\scalerel*{
\begin{tikzpicture}[yscale=-1,transform shape]
\pic{orcidlogo};
\end{tikzpicture}
}{|}}}}
\makeatother
\begin{document}
\title{Noether and Mei symmetries in static spherically symmetric quadratic gravity: variational consistency, constraints, and conserved curvature flux}
\author{G.~G.~N.~Nashed$^{1,2}$}\email{nashed@bue.edu.eg}
\author{A.~Eid$^{3}$}\email{amaid@imamu.edu.sa}
\author{Kazuharu Bamba$^{4}$}
\email{bamba@sss.fukushima-u.ac.jp}
\affiliation{Centre for Theoretical Physics, The British University in Egypt, P.O. Box 43, El Sherouk City, Cairo 11837, Egypt\\$^{2}$Centre for Space Research (CSR), North-West University, Potchefstroom 2520, South Africa\\$^{3}$Department of Physics, College of Science, Imam Mohammad Ibn Saud Islamic University (IMSIU), Riyadh, Kingdom of Saudi Arabia\\$^{4}$Faculty of Symbiotic Systems Science, Fukushima University, Fukushima 960-1296, Japan}

\begin{abstract}
The symmetry content of static spherical $f(R)$ gravity is reconsidered for the pure quadratic model. We first derive the radial action without prematurely eliminating the equation carried by the radial metric variable. The Schwarzschild-type gauge may then be imposed while retaining its associated gravitational constraint. For the resulting system, radial translations and one combined scaling are exact off-shell Noether symmetries. The scaling charge reduces, on the constraint surface, to a conserved radial flux of the scalar curvature. The same flux follows independently from the trace of the four-dimensional field equations. Noether invariance, the strong point-Mei criterion, and Lie invariance of the Euler--Lagrange system are examined separately. Within the full polynomial point ansatz of total degree at most two, the Lie algebra contains precisely three independent generators. The flux also separates the solution space into a nonzero constant-curvature Einstein sector, a degenerate scalar-flat sector, and a dynamical-curvature sector. In particular, the scalar-flat sector permits a Reissner--Nordstr\"om-form metric, although its inverse-square coefficient has no electromagnetic meaning in the absence of a Maxwell field.
\end{abstract}

\maketitle

\section{Introduction}
Modified theories of gravity have been widely considered to describe gravitational phenomena that may not be fully explained within General Relativity. A well-known example is metric $f(R)$ gravity, in which the Einstein--Hilbert term is generalized by allowing the gravitational Lagrangian to depend nonlinearly on the Ricci scalar. Apart from its physical applications, this theory provides a convenient framework for studying the relation between spacetime symmetries, reduced dynamical systems, and conserved quantities. In particular, the existence of a variational symmetry may simplify the field equations and lead to a first integral that can be used in the search for exact solutions \cite{SotiriouFaraoni2010,Bahamonde2019,Bajardi2023,Capozziello2007}.

Several aspects of metric and metric-affine $f(R)$ gravity have already been investigated. These include the mathematical formulation of the theory, its scalar--tensor representation, and its applications to cosmology and gravitational phenomenology
\cite{DeFelice:2010aj,Nojiri:2010wj,Nojiri:2017ncd,Clifton:2011jh,
Capozziello:2011et,Nojiri:2006ri,Faraoni:2008mf,
Sotiriou:2006hs,Sotiriou:2006mu,Whitt:1984pd,Chiba:2003ir}.
However, when a spacetime symmetry is imposed at the level of the action, some care is required. The introduction of an auxiliary curvature variable, the treatment of boundary contributions, and the fixing of a coordinate gauge may change the reduced variational problem if these operations are not performed in the proper order.

The static spherically symmetric configuration is particularly suitable for examining this question. In this geometry, the four-dimensional gravitational action can be reduced to a one-dimensional system whose independent variable is the radial coordinate. Noether and Mei symmetries of such a system have recently been discussed for the quadratic $f(R)$ model, and a number of generators and corresponding currents have been presented \cite{Dabash2025,Paliathanasis2026,DabashReply2026}. Nevertheless, the regularity of the Legendre transformation does not by itself establish the variational equivalence of two radial Lagrangians. A term that contains no radial derivatives does not enter the Hessian, but it may still appear in the Euler--Lagrange equations through its derivatives with respect to the configuration variables. Therefore, such a term cannot be omitted solely because it is independent of the generalized velocities.

Noether and Noether-gauge symmetries have been applied to both metric and Palatini formulations of $f(R)$ gravity. In several cases, the symmetry conditions restrict the admissible form of $f(R)$ and provide conserved quantities that help integrate the reduced equations
\cite{Capozziello:2008ch,Martin-Moruno:2008qpc,
Paliathanasis:2011jq,Capozziello:2012iea,Hussain:2011wa,
Vakili:2008uj,Kucuakca:2011np}.
Static spherical solutions have also been obtained in $f(R)$ gravity and in other higher-derivative theories by using analytical and numerical approaches
\cite{Multamaki:2006zb,delaCruz-Dombriz:2009pzc,
Sebastiani:2010kv,Nashed:2019uyi}.
These results motivate a careful examination of the radial variational formulation before its symmetry algebra is determined.

The origin of the present investigation is the recent Comment--Reply discussion concerning the canonical reduction used in the previous symmetry analysis. Paliathanasis argued that the adopted minisuperspace Lagrangian does not contain an essential curvature-dependent contribution. In their Reply, Dabash, Emam, and Sch"oppner maintained that their gauge-fixed canonical formulation is consistent \cite{Paliathanasis2026,DabashReply2026}. Instead of beginning with either conclusion, we derive the radial Lagrangian from the four-dimensional action and retain the equation associated with the radial metric variable before the gauge condition is imposed. The proposed transformations can then be tested directly against the resulting Lagrangian and equations of motion.

The present study is devoted to pure quadratic gravity. We first examine which of the previously considered transformations satisfy the Noether condition when the complete radial Lagrangian is used. We then compare these transformations with the strong point-Mei criterion and with the Lie invariance condition for the Euler--Lagrange equations. These three conditions are treated separately because a transformation that satisfies one of them need not satisfy the others. We also investigate the covariant origin of the first integral obtained from the Noether symmetry. The choice $f(R)=R^2$ is useful for this purpose because the theory possesses a simple scaling behavior and its trace equation takes a particularly direct form.

Quadratic gravitational theories have been studied since the early investigations of nonlinear curvature actions and renormalizable gravity, as well as in connection with the Starobinsky inflationary model
\cite{Buchdahl:1970ynr,Stelle:1976gc,Starobinsky:1980te}.
Their static spherical sector is not restricted to the Schwarzschild solution. It includes Einstein spaces, non-Schwarzschild black holes, and other regular or singular geometries
\cite{Lu:2015psa,Lu:2015cqa,Holdom:2016xfn,Pravda:2020zno,
Donoghue:2021cza,Hernandez-Lorenzo:2020aie,Podolsky:2018pfe,
Kokkotas:2017zwt,Bueno:2017sui,Berej:2006cc,Pravda:2016fue,
Svarc:2018coe,Nelson:2010ig}.
In the pure $R^2$ theory, the scalar-flat sector is also degenerate because both the Lagrangian and its first derivative vanish at $R=0$. This property plays an important role in the interpretation of the solution branches discussed below.

It should be stressed that verifying several proposed generators is different from classifying the symmetry algebra. Substitution of a particular generator into a prolongation condition proves only whether that generator is admitted. It does not exclude additional transformations. For this reason, we supplement the direct tests by solving the Lie determining equations for the complete polynomial point ansatz of total degree not exceeding two. The resulting statement is complete within this polynomial class only. Non-polynomial point transformations and generalized symmetries whose coefficients depend on derivatives are outside the scope of that classification.

The arrangement of this paper is as follows. In Sec.~\ref{sec:reduction}, we derive the static spherical radial Lagrangian and recover the gravitational constraint associated with the radial gauge. The Noether symmetries, strong point-Mei tests, and polynomial Lie-point classification are presented in Sec.~\ref{sec:noether}. In Sec.~\ref{sec:covariant}, the first integral is obtained independently from the covariant trace equation, and the corresponding solution sectors are discussed. A comparison with the previously proposed construction is given in Sec.~\ref{sec:comparison}. The local and asymptotic consequences of the curvature-flux relation are examined in Sec.~\ref{sec:fluxconsequences}. Section~\ref{sec:scope} states the range of the classification and the remaining open problems, while Sec.~\ref{sec:conclusion} contains the conclusions. The symbolic identities used to verify the calculations are collected in Appendix~\ref{app:verification}.

\section{Variationally Consistent Spherical Reduction}
\label{sec:reduction}

We begin with the vacuum metric $f(R)$ action
\begin{equation}
S[g]=\int \dd^4x\,\sqrt{-g}\,f(R),
\label{eq:action}
\end{equation}
where $R$ denotes the Ricci scalar. Metric variation gives
\begin{equation}
f_R R_{\mu\nu}
-\frac{1}{2}f g_{\mu\nu}
+\left(
g_{\mu\nu}\Box-\nabla_\mu\nabla_\nu
\right)f_R=0,
\label{eq:field}
\end{equation}
where
\begin{equation}
f_R\equiv\frac{\dd f}{\dd R}.
\label{eq:fRdef}
\end{equation}

Before fixing the radial coordinate, take the static spherical line element in the form
\begin{equation}
\dd s^2=-A(r)\dd t^2+B(r)\dd r^2+N(r)\dd\Omega^2,
\qquad
\dd\Omega^2=\dd\theta^2+\sin^2\theta\,\dd\phi^2.
\label{eq:generalmetric}
\end{equation}
The three functions $A$, $B$, and $N$ are kept independent in the variation. Imposing $AB=1$ beforehand would eliminate both a radial metric variable and its equation. We identify the constraint first and only then choose the Schwarzschild-type gauge
\begin{equation}
A=H,
\qquad
B=H^{-1},
\label{eq:gauge}
\end{equation}
which gives
\begin{equation}
\dd s^2=-H(r)\dd t^2+\frac{\dd r^2}{H(r)}
+N(r)\dd\Omega^2,
\qquad
\sqrt{-g}=N\sin\theta.
\label{eq:metric}
\end{equation}

We next introduce $R$ as an independent minisuperspace coordinate, enforce its definition with a Lagrange multiplier, and integrate the genuine second-radial-derivative terms by parts. The unfixed first-order radial Lagrangian is
\begin{equation}
L_{\mathrm{u}}=
\sqrt{AB}\left\{
N\left(f-Rf_R\right)
+2f_R
+f_R\left[
\frac{(N')^2}{2BN}
+\frac{A'N'}{AB}
\right]
+f_{RR}R'\left[
\frac{2N'}{B}
+\frac{NA'}{AB}
\right]
\right\}.
\label{eq:unfixedL}
\end{equation}
Since Eq.~\eqref{eq:unfixedL} contains no $B'$, its $B$ variation is a constraint rather than a radial evolution equation. Applying Eq.~\eqref{eq:gauge} after this variation gives
\begin{equation}
L_f=
N(f-Rf_R)
+f_R\left[
\frac{H(N')^2}{2N}
+H'N'
+2
\right]
+f_{RR}R'
\left(
2HN'+NH'
\right).
\label{eq:Lf}
\end{equation}

The last expression contains the velocity-independent term $2f_R$. It makes no contribution to the reduced Hessian, but for $f_{RR}\neq0$ it still enters the dynamics because
\begin{equation}
\frac{\partial(2f_R)}{\partial R}=2f_{RR}.
\label{eq:potentialvariation}
\end{equation}
It cannot therefore be discarded merely because it contains no radial velocity.

For the pure quadratic model,
\begin{equation}
f(R)=R^2,
\label{eq:quadraticmodel}
\end{equation}
for which
\begin{equation}
f_R=2R,
\qquad
f_{RR}=2.
\label{eq:quadraticderivatives}
\end{equation}
Eq.~\eqref{eq:Lf} becomes
\begin{equation}
L=
\frac{RH}{N}(N')^2
+2RH'N'
+4HR'N'
+2NR'H'
+4R
-NR^2.
\label{eq:L}
\end{equation}
All symmetry calculations below start from Eq.~\eqref{eq:L}.

\subsection{Quadratic equations, momenta, and radial energy}
\label{sec:eom}

Using the convention
\begin{equation}
\mathcal{E}_q(L)=
\frac{\dd}{\dd r}
\left(
\frac{\partial L}{\partial q'}
\right)-
\frac{\partial L}{\partial q},
\end{equation}
the equations obtained by variation with respect to $H$, $N$, and $R$ can be written, after multiplication by nonzero powers of $N$, as
\begin{equation}
2N^2R''+2NN''R-(N')^2R=0,
\label{eq:EH}
\end{equation}
\begin{align}
0={}&4HN^2R''+2HNN'R'+2HNN''R-H(N')^2R+4H'N^2R'+2H'NN'R+2H''N^2R+N^2R^2,
\label{eq:EN}
\end{align}
and
\begin{equation}
4HNN''-H(N')^2+4H'NN'+2H''N^2+2N^2R-4N=0.
\label{eq:ER}
\end{equation}
Equation~\eqref{eq:ER} gives
\begin{equation}
R=
\frac{
4N-4HNN''+H(N')^2-4H'NN'-2H''N^2
}{
2N^2
},
\label{eq:Rgeometry}
\end{equation}
which agrees with the Ricci scalar calculated directly from Eq.~\eqref{eq:metric}. This checks the auxiliary-curvature reduction.

The canonical momenta are
\begin{align}
p_H&=2RN'+2NR',
\label{eq:pH}\\
p_N&=\frac{2RH}{N}N'+2RH'+4HR',
\label{eq:pN}\\
p_R&=4HN'+2NH'.
\label{eq:pR}
\end{align}
The determinant of the velocity Hessian is
\begin{equation}
\det\left(
\frac{\partial^2L}{\partial q'^i\partial q'^j}
\right)=
24HNR,
\label{eq:hessian}
\end{equation}
Thus the gauge-fixed Legendre map is regular wherever $HNR\neq0$. This statement concerns the reduced mechanical problem and does not remove the gravitational constraint inherited from the unfixed system.

The radial energy function is
\begin{equation}
{
E_L=
NR^2-4R
+\frac{RH}{N}(N')^2
+2RH'N'
+4HR'N'
+2NR'H'
}.
\label{eq:energy}
\end{equation}
The constraint inherited from the unfixed system can now be derived
explicitly. Before specializing to $f(R)=R^2$, let $E_f$ denote the
energy function associated with the gauge-fixed Lagrangian in
Eq.~\eqref{eq:Lf}. Direct differentiation of the unfixed Lagrangian
in Eq.~\eqref{eq:unfixedL} with respect to the lapse-like variable
$B$, followed by imposing $A=H$ and $B=H^{-1}$, gives
\begin{equation}
\left.
\frac{\partial L_{\mathrm{u}}}{\partial B}
\right|_{A=H,\,B=H^{-1}}
=
-\frac{H}{2}E_f,
\label{eq:Bconstraintidentity}
\end{equation}
where
\begin{equation}
E_f
=
N(Rf_R-f)-2f_R
+
f_R
\left[
\frac{H(N')^2}{2N}+H'N'
\right]
+
f_{RR}R'
\left(
2HN'+NH'
\right).
\label{eq:generalenergy}
\end{equation}
Since $B'$ does not appear in $L_{\mathrm{u}}$, its Euler--Lagrange
equation, with the convention adopted above, is
\begin{equation}
\mathcal{E}_B(L_{\mathrm{u}})
=
-\frac{\partial L_{\mathrm{u}}}{\partial B}
=
0.
\label{eq:BELequation}
\end{equation}
Combining Eqs.~\eqref{eq:Bconstraintidentity} and
\eqref{eq:BELequation} therefore yields, on every regular patch
with $H\neq0$,
\begin{equation}
E_f=0.
\label{eq:generalconstraint}
\end{equation}
For $f(R)=R^2$, Eq.~\eqref{eq:generalenergy} reduces exactly to
the radial energy function $E_L$ in Eq.~\eqref{eq:energy}. Hence
the constraint inherited from the unfixed variational system is
\begin{equation}
{E_L=0.}
\label{eq:constraint}
\end{equation}
Thus, $E_L=0$ is not imposed as an independent zero-energy condition
on the gauge-fixed mechanical system; rather, it is the metric
equation associated with $B$ that is removed upon imposing the radial
gauge $AB=1$. This also explains how the gauge-fixed Hessian can be
nondegenerate while the parent generally covariant system still
supplies a separate gravitational constraint.

\section{Verified Noether Symmetries and First Integral}\label{sec:noether}

Consider the point generator
\begin{equation}
 X=\xi\partial_r+\eta^H\partial_H+\eta^N\partial_N+\eta^R\partial_R,
\end{equation}
with first prolongation
\begin{equation}
 X^{[1]}=X+\sum_{q=H,N,R}\left(D_r\eta^q-q'D_r\xi\right)\partial_{q'},
\end{equation}
The Noether condition, including a gauge function $K$, reads
\begin{equation}
 X^{[1]}L+L D_r\xi=D_rK.
 \label{eq:noethercondition}
\end{equation}

Direct substitution verifies
\begin{equation}
 X_1=\partial_r,
 \label{eq:X1}
\end{equation}
and
\begin{equation}
 {X_N=r\partial_r+H\partial_H+N\partial_N-R\partial_R}
 \label{eq:XN}
\end{equation}
as exact off-shell Noether symmetries with constant $K$.  In particular,
\begin{equation}
 X_N^{[1]}L+L=0.
 \label{eq:XNresidual}
\end{equation}

For comparison, test separately the two scalings reported for the reduced $R^2$ model in Refs.~\cite{Dabash2025,Paliathanasis2026,DabashReply2026}:
\begin{align}
 Y_2=r\partial_r+H\partial_H,\ \quad 
 Y_3=H\partial_H+N\partial_N-R\partial_R.
\end{align}
For the Lagrangian \eqref{eq:L}, direct calculation gives
\begin{align}
 Y_2^{[1]}L+L&=4R-NR^2,\label{eq:Y2res}\\
 Y_3^{[1]}L&=E_L.\label{eq:Y3res}
\end{align}
Thus $Y_2$ fails the off-shell Noether test for Eq.~\eqref{eq:L}. The residual of $Y_3$ vanishes only on the gravitational constraint surface and is not an off-shell variational identity. Their sum, however, is exactly the scaling in Eq.~\eqref{eq:XN}.

For $X_N$, the Noether invariant may be written
\begin{equation}
 I_N=Hp_H+Np_N-Rp_R-rE_L.
 \label{eq:INdef}
\end{equation}
Equations~\eqref{eq:pH}--\eqref{eq:pR} give the identity
\begin{equation}
 Hp_H+Np_N-Rp_R=6HNR',
 \label{eq:momentumidentity}
\end{equation}
and therefore
\begin{equation}
 {I_N=6HNR'-rE_L.}
 \label{eq:IN}
\end{equation}
On the gravitational constraint surface,
\begin{equation}
 {HNR'=C_R.}
 \label{eq:flux}
\end{equation}
The verified generators satisfy $[X_1,X_N]=X_1$. These calculations establish a two-generator sector; completeness in the unrestricted nonlinear point class would require the full Noether determining system.

\subsection{Strong point-Mei tests}\label{sec:mei}

Following the form-invariance criterion used in the Mei literature \cite{Mei2000,Zhang2021,Asghar2022}, we call a point transformation a \emph{strong point-Mei symmetry} here when
\begin{equation}
 \mathcal E_i\!\left(X^{[1]}L\right)=0,
 \qquad i=H,N,R.
 \label{eq:mei}
\end{equation}
This criterion is distinct from the Noether identity in Eq.~\eqref{eq:noethercondition}.

Direct first-prolongation calculations verify
\begin{equation}
 M_1=\partial_r,
 \qquad
 M_2=\frac12r\partial_r+H\partial_H.
 \label{eq:Meigenerators}
\end{equation}
For the second generator one finds the stronger identity
\begin{equation}
 M_2^{[1]}L=0,
 \label{eq:M2strong}
\end{equation}
whereas
\begin{equation}
 M_2^{[1]}L+L D_r\!\left(\frac r2\right)=\frac12L.
 \label{eq:M2notNoether}
\end{equation}
Thus, $M_2$ passes the strong Mei test but not the ordinary Noether test for Eq.~\eqref{eq:L}.

References~\cite{Dabash2025,Paliathanasis2026,DabashReply2026} obtain further expressions after imposing an affine point ansatz for $\xi$ and $\eta^i$. Some of the later formulas depend on derivatives of the dynamical variables. Such objects may define generalized symmetries, but they do not belong to the stated point class unless the problem is reformulated on an appropriate jet space. We therefore draw no conclusion about them from the point calculation and keep the verified point sector separate from a possible generalized Mei analysis.

\subsection{Polynomial Lie-point classification through quadratic order}\label{sec:lie}

For a point symmetry of the second-order system \eqref{eq:EH}--\eqref{eq:ER}, the appropriate invariance condition is
\begin{equation}
 X^{[2]}(E_i)\big|_{E_H=E_N=E_R=0}=0,
 \qquad i=H,N,R,
 \label{eq:lietest}
\end{equation}
where
\begin{equation}
 X=\xi(r,H,N,R)\partial_r+\eta^H\partial_H+\eta^N\partial_N+\eta^R\partial_R.
 \label{eq:generalLie}
\end{equation}
To obtain a classification rather than a list of successful tests, we solve the determining equations in the full polynomial point class of total degree at most two. Let
\begin{equation}
 m_A\in\{1,r,H,N,R,r^2,rH,rN,rR,H^2,HN,HR,N^2,NR,R^2\}
 \label{eq:quadbasis}
\end{equation}
be the 15 monomials of degree $\leq2$.  We take
\begin{equation}
 \xi=\sum_{A=1}^{15}a_A m_A,
 \quad
 \eta^H=\sum_{A=1}^{15}b_A m_A,
 \quad
 \eta^N=\sum_{A=1}^{15}c_A m_A,
 \quad
 \eta^R=\sum_{A=1}^{15}d_A m_A,
 \label{eq:quadansatz}
\end{equation}
so that the ansatz contains 60 independent constants.

The second prolongation is formed, $(H'',N'',R'')$ are eliminated with Eqs.~\eqref{eq:EH}--\eqref{eq:ER}, and the result is split over independent monomials in $(r,H,N,R,H',N',R')$. There are 893 coefficient equations before duplicates are removed and 495 distinct linear conditions afterward. Their coefficient matrix has
\begin{equation}
 \operatorname{rank}=57,
 \qquad
 \operatorname{nullity}=60-57=3.
 \label{eq:lrank}
\end{equation}
A convenient nullspace basis is
\begin{align}
 L_1&=\partial_r,\qquad  L_2=\frac12r\partial_r+H\partial_H,\qquad
 L_4=-\frac12r\partial_r-N\partial_N+R\partial_R.
 \label{eq:LieNullBasis}
\end{align}
Equivalently, since $L_3=L_2-L_4$, one may use the physically transparent basis
\begin{align}
 L_1=\partial_r,\ \quad 
 L_2=\frac12r\partial_r+H\partial_H,\ \quad 
 L_3=r\partial_r+H\partial_H+N\partial_N-R\partial_R.
 \label{eq:LieBasis}
\end{align}
The above normal-form calculation is carried out on the regular sector \(HNR\neq0\), where Eqs.~(14)--(16) can be solved for the highest derivatives \((H'',N'',R'')\). This restriction concerns the implementation of the Lie determining equations and does not exclude degenerate solution sectors of the original field equations. In particular, the scalar-flat branch (R=0), discussed separately in Secs.~IV A and VI D, remains a valid sector of pure \(R^2\) gravity but lies outside the regular normal-form domain used in the polynomial Lie classification.

Accordingly, the point-symmetry algebra within this polynomial class is
\begin{equation}
 {
 \mathfrak g_{\mathrm {Lie}}^{(\deg\leq2)}
 =\operatorname{span}\{L_1,L_2,L_3\}.
 }
 \label{eq:LieQuadraticClass}
\end{equation}
In particular, there are no additional constant, linear, or quadratic polynomial Lie-point generators.

Their nonvanishing brackets are
\begin{equation}
 [L_1,L_2]=\frac12L_1,
 \qquad
 [L_1,L_3]=L_1,
 \qquad
 [L_2,L_3]=0.
 \label{eq:brackets}
\end{equation}
With $D=2L_2$ and $S=L_3-2L_2$, one has $[L_1,D]=L_1$ and $S$ central, so the verified quadratic polynomial algebra is isomorphic to
\begin{equation}
 {\mathfrak{aff}(1)\oplus\mathbb R.}
 \label{eq:LieStructure}
\end{equation}
This goes beyond an affine ansatz, but it does not exclude rational, logarithmic, exponential, or other non-polynomial point generators. Equation~\eqref{eq:LieQuadraticClass} is therefore not asserted to be the unrestricted local point algebra.

\section{Covariant interpretation: conserved curvature flux}\label{sec:covariant}

We now determine the covariant meaning of the conserved quantity. The
relation between variational symmetries, constraints, covariant phase space,
and Noether charges in generally covariant theories is well established
\cite{Lee:1990nz,Wald:1993nt,Iyer:1994ys}. Taking the trace of
Eq.~\eqref{eq:field}, we obtain
\begin{equation}
 3\Box f_R+Rf_R-2f=0.
 \label{eq:trace}
\end{equation}
For $f(R)=R^2$ this reduces to
\begin{equation}
 {\Box R=0.}
 \label{eq:boxR}
\end{equation}
For Eq.~\eqref{eq:metric}, $R=R(r)$ and
\begin{equation}
 \Box R=\frac{1}{\sqrt{-g}}\partial_r\!\left(\sqrt{-g}\,g^{rr}R'\right)
 =\frac1N\frac{\dd}{\dd r}(HNR').
 \label{eq:boxradial}
\end{equation}
The trace equation therefore gives
\begin{equation}
 \frac{\dd}{\dd r}(HNR')=0,
\end{equation}
and hence Eq.~\eqref{eq:flux}. Thus the constrained Noether charge and the four-dimensional trace equation independently yield the same radial constant.

For the unfixed areal-radius metric
\begin{equation}
 \dd s^2=-A(r)\dd t^2+B(r)\dd r^2+r^2\dd\Omega^2,
\end{equation}
the corresponding flux is
\begin{equation}
 {r^2\sqrt{\frac AB}\,R'=C_R.}
 \label{eq:unfixedflux}
\end{equation}

\subsection{Solution sectors}\label{sec:solutions}

The constant $C_R$ divides the solution space naturally. If $C_R=0$, Eq.~\eqref{eq:flux} gives $R'=0$ on a regular region with $HN\neq0$, and therefore $R=R_0$.

For $R_0\neq0$, Eq.~\eqref{eq:field} becomes
\begin{equation}
 R_{\mu\nu}=\frac{R_0}{4}g_{\mu\nu},
\end{equation}
and the areal-radius solution is the Schwarzschild--(anti-)de Sitter family
\begin{equation}
 H(r)=1-\frac{2M}{r}-\frac{R_0}{12}r^2
 =1-\frac{2M}{r}-\frac{\Lambda r^2}{3},
 \qquad R_0=4\Lambda.
 \label{eq:SdS}
\end{equation}
Here the simultaneous appearance of $N=r^2$ and $AB=1$ should not be interpreted as the imposition of two independent radial gauge conditions on the general system. In the nonzero constant-curvature branch, the field equations reduce to the Einstein-space condition $R_{\mu\nu}=(R_0/4)g_{\mu\nu}$, whose static, spherically symmetric solution can be expressed in areal-radius coordinates in the Schwarzschild--(anti-)de Sitter form~\eqref{eq:SdS}. The restriction against imposing both conditions independently applies to the generic dynamical-curvature sector.

The branch $R_0=0$ is degenerate in pure quadratic gravity because
\begin{equation}
f(0)=0, \qquad f_R(0)=0.
\end{equation}
Consequently, the vacuum field equations do not require
$R_{\mu\nu}=0$ in this sector, and scalar flatness alone does not
uniquely determine the metric. In particular, an admissible
static, spherically symmetric scalar-flat solution is
\begin{equation}
H(r)=1-\frac{2M}{r}+\frac{C}{r^2}.
\label{eq:RNform}
\end{equation}
This metric has the Reissner--Nordstr\"om functional form, but in
the pure gravitational theory considered here the integration
constant $C$ is not, by itself, an electromagnetic charge. Such an
interpretation requires an additional matter sector, such as
Maxwell electrodynamics.

When $C_R\neq0$, the curvature is dynamical:
\begin{equation}
 R'=\frac{C_R}{HN}.
\end{equation}
This branch deserves a separate global analysis.  In particular, one should not generically impose both the radial gauge $AB=1$ and the areal-radius choice $N=r^2$ as independent coordinate conditions.

\section{Comparison with the previously proposed symmetry construction}\label{sec:comparison}

References~\cite{Dabash2025,Paliathanasis2026,DabashReply2026} study the same static spherical setting. They construct a canonical $f(R)$ Lagrangian and, for $R^2$, report radial and scaling Noether generators together with eight Mei generators and associated currents. The present comparison cannot be settled from the Hessian alone, because the reduced potential contributes both to the Euler--Lagrange equations and to the velocity-independent part of the symmetry conditions.

The later Comment by Paliathanasis identifies the variational reduction as the central problem and argues that the resulting conservation-law analysis is consequently unreliable \cite{Paliathanasis2026}.  The authors' Reply rejects that interpretation and emphasizes the legitimacy of fixing the radial gauge before carrying out the canonical symmetry analysis \cite{DabashReply2026}.  These two published positions should be read together with the original article.  Our purpose here is narrower: we ask whether the displayed one-dimensional Lagrangian, with or without the disputed curvature-dependent term, yields the stated Euler--Lagrange equations and symmetry residuals.

Our reconstruction retains the $2f_R$ term in Eq.~\eqref{eq:Lf}, equal to $4R$ for the quadratic model. Equations~\eqref{eq:Y2res} and \eqref{eq:Y3res} then give a direct diagnostic: the residuals of $Y_2$ and $Y_3$ are $4R-NR^2$ and $E_L$, respectively, whereas their sum $X_N=Y_2+Y_3$ is an off-shell variational symmetry.

\subsection{A direct variational diagnostic}

The issue can be stated without reference to the Hessian. Let
\begin{equation}
 L_{\mathrm {full}}=L_{\mathrm { red}}+2f_R .
\end{equation}
Since $2f_R$ contains no radial velocities, the two Lagrangians have the same velocity Hessian. Nevertheless,
\begin{equation}
 \frac{\partial}{\partial R}(2f_R)=2f_{RR},
\end{equation}
which is nonzero for a nonlinear theory.  For $f(R)=R^2$ this difference is $4R$ and contributes a constant $4$ to the $R$-variation.  Thus equality of the Legendre Hessians is not sufficient to establish equality of the variational dynamics.  The relevant test is whether the difference is a total radial derivative.  A function of $R$ alone cannot equal $D_rF(r,H,N,R)$ for arbitrary paths without generating velocity terms unless its $R$ dependence is trivial.  This observation explains why the disputed term must be tracked explicitly when the symmetry determining equations are constructed.

Gauge fixing introduces a second, related distinction. The choice $AB=1$ is consistent only when the equation of the eliminated radial variable is retained. In the reduced mechanics it appears as $E_L=0$. An identity valid only after imposing this equation is consequently weaker than an off-shell Noether identity. This is the difference between $Y_3$ and the exact combined scaling $X_N$.

The comparison of the verified sectors is summarized in Table~\ref{tab:symmetry}.
\begin{table}[t]
\caption{Directly verified symmetry properties for the corrected quadratic Lagrangian. ``Yes'' means the corresponding prolongation test vanishes in the sense stated in the text. The table is not a completeness theorem for unrestricted nonlinear or generalized generators.}
\label{tab:symmetry}
\begin{ruledtabular}
\begin{tabular}{lccc}
Generator & Noether & Strong Mei & Lie EOM\\
\hline
$\partial_r$ & yes & yes & yes\\
$\tfrac12r\partial_r+H\partial_H$ & no & yes & yes\\
$r\partial_r+H\partial_H+N\partial_N-R\partial_R$ & yes & no & yes\\
\end{tabular}
\end{ruledtabular}
\end{table}

A derivative-dependent expression likewise cannot be inferred from a point ansatz without changing the symmetry problem. Such candidates require generalized prolongations on the relevant jet space and are not included in the point-sector statements made here.

\subsection{Scaling weights and invariant variables}\label{sec:scalingreduction}

The exact Noether generator \eqref{eq:XN} can be used not only to construct a first integral but also to organize the reduced variables into scale-invariant combinations.  Its characteristic equations are
\begin{equation}
 \frac{\dd r}{r}=\frac{\dd H}{H}=\frac{\dd N}{N}=-\frac{\dd R}{R}.
 \label{eq:characteristics}
\end{equation}
Thus one may assign the weights
\begin{equation}
 w(r)=w(H)=w(N)=1,\qquad w(R)=-1,
 \label{eq:weights}
\end{equation}
under the one-parameter transformation generated by $X_N$.  Three independent zeroth-order invariants are, for example,
\begin{equation}
 u=\frac{H}{r},\qquad v=\frac{N}{r},\qquad w=rR.
 \label{eq:invariants}
\end{equation}
Equivalently, introducing the logarithmic radial coordinate $s=\ln r$ converts the scaling part of $X_N$ into a translation in $s$ after the dependent variables are rescaled according to Eq.~\eqref{eq:weights}.  This provides a systematic route to symmetry reduction of the field equations without imposing the areal-radius condition $N=r^2$ on top of the gauge $AB=1$.

This formulation separates the coordinate rescaling generated by $r\partial_r$ from the compensating transformations of $(H,N,R)$. It also turns self-similar configurations into fixed points of the associated autonomous system. Equation~\eqref{eq:invariants} thus provides a starting point for a phase-space analysis of the $C_R\neq0$ sector, without assuming that every fixed point describes a regular black hole.

The Noether charge itself has a transparent scaling interpretation.  On the constraint surface it is proportional to
\begin{equation}
 \mathcal F_R=HNR',
 \label{eq:FRdef}
\end{equation}
which is invariant along the radial evolution by Eq.~\eqref{eq:flux}.  Therefore the symmetry reduction and the covariant trace equation select the same one-dimensional invariant of the flow.

\section{Consequences of the curvature-flux law}\label{sec:fluxconsequences}

Even when the remaining equations cannot be integrated in closed form, Eq.~\eqref{eq:flux} imposes local restrictions. On a regular patch with $HN\neq0$,
\begin{equation}
 R'=\frac{C_R}{HN}.
 \label{eq:Rprimeflux}
\end{equation}
Consequently, a nonzero $C_R$ forbids a stationary point of $R$ inside such a patch.  The sign of $R'$ is fixed by the sign of $C_R/(HN)$ until a zero or singularity of the denominator is encountered.  Thus $R$ is locally monotonic on every connected regular region in which $H$ and $N$ retain their signs.

Hence a proposed $C_R\neq0$ geometry with $R'(r_*)=0$ and $H(r_*)N(r_*)\neq0$ is incompatible with the trace equation. Conversely, $C_R=0$ enforces constant scalar curvature on every regular connected region.

A horizon requires additional care. In the gauge~\eqref{eq:metric}, a Killing horizon is typically located at \(H(r_h)=0\). If \(C_R\neq 0\) and \(N(r_h)\) is finite and nonzero, Eq.~\eqref{eq:Rprimeflux} implies that \(R'\) cannot remain finite in these coordinates unless its near-horizon behavior compensates for the vanishing of \(H\). This is only a local diagnostic, since \(R'\) is coordinate dependent; regularity must ultimately be assessed in a horizon-regular coordinate system using curvature invariants. Nevertheless, the result indicates that the nonzero-flux branch has a qualitatively different near-horizon structure from the constant-curvature branch and cannot generally be obtained from it through a small algebraic deformation.

The unfixed areal-radius form \eqref{eq:unfixedflux} is particularly convenient for this purpose:
\begin{equation}
 R'=\frac{C_R}{r^2}\sqrt{\frac{B}{A}}.
 \label{eq:arealRprime}
\end{equation}
It keeps $A$ and $B$ independent and therefore avoids imposing two radial coordinate conditions simultaneously.  Equations \eqref{eq:arealRprime} and the remaining independent metric field equations form a natural system for numerical shooting or asymptotic analysis of the $C_R\neq0$ sector.

\subsection{Asymptotic implication}

Suppose an asymptotic region admits $A\to A_\infty\neq0$ and $B\to B_\infty\neq0$.  Equation~\eqref{eq:arealRprime} then gives, at leading order,
\begin{equation}
 R'\sim \frac{C_R}{r^2}\sqrt{\frac{B_\infty}{A_\infty}},
 \end{equation}
and hence
\begin{equation}
 R(r)=R_\infty-\frac{C_R}{r}\sqrt{\frac{B_\infty}{A_\infty}}+o(r^{-1}),
 \label{eq:Rasymptotic}
\end{equation}
provided the assumed limits and the asymptotic expansion exist.  Thus the conserved curvature flux controls the leading $1/r$ scalar-curvature tail in such a region.  Equation~\eqref{eq:Rasymptotic} is a conditional asymptotic consequence of the trace equation, not a complete solution of the metric equations.

\subsection{Constraint-shell versus off-shell symmetry}\label{sec:offshell}

The residual \eqref{eq:Y3res} illustrates a distinction that is often obscured after gauge fixing.  A transformation satisfying
\begin{equation}
 X^{[1]}L+LD_r\xi=0
\end{equation}
identically on the jet space is an off-shell variational symmetry.  By contrast, a transformation whose residual is proportional to a constraint is a symmetry only after restriction to the constraint surface.  In the present case
\begin{equation}
 Y_3^{[1]}L=E_L,
\end{equation}
so the vanishing of this residual requires Eq.~\eqref{eq:constraint}.  This is weaker than the identity \eqref{eq:XNresidual} obeyed by $X_N$.

This distinction also affects conserved quantities.  Noether's theorem applies directly to an off-shell variational identity and produces a first integral on solutions of the Euler--Lagrange equations.  A constraint-shell relation can still be dynamically useful, but its status depends on the constraint structure inherited from the parent gravitational action.  For this reason the present paper keeps the statements ``off-shell Noether symmetry'' and ``constraint-shell invariance'' separate throughout.

The same logic explains why regularity of the gauge-fixed Hessian does not imply absence of a gravitational constraint.  Gauge fixing can turn a singular generally covariant variational problem into a regular mechanical one while simultaneously removing an equation that must be restored from the unfixed system.  The determinant \eqref{eq:hessian} therefore characterizes the reduced Legendre map, whereas Eq.~\eqref{eq:constraint} carries information inherited from radial diffeomorphism invariance.

\subsection{Point and generalized Mei transformations}\label{sec:generalizedmei}

The original analysis motivating the present reconstruction states a point-generator ansatz and reports eight Mei generators for the quadratic model.  Its stated objective is to solve an overdetermined PDE system for the generator components and construct currents; the published paper explicitly describes the eight generators as independent \cite{Dabash2025,Paliathanasis2026,DabashReply2026}.  The point/generalized distinction is therefore not merely terminological.

For a point transformation, the coefficients $\xi$ and $\eta^i$ depend on the independent and dependent variables but not on derivatives.  If instead
\begin{equation}
 \eta^i=\eta^i(r,H,N,R,H',N',R',\ldots),
\end{equation}
then the transformation is generalized and its prolongation must be computed on the corresponding jet space.  Derivative-dependent coefficients cannot be inserted consistently into point determining equations after those equations have been derived under the assumption that the coefficients are derivative independent.

This observation does not establish that a derivative-dependent candidate is false.  It establishes that the candidate belongs to a different symmetry problem.  A complete follow-up analysis should therefore proceed in two stages: first solve the unrestricted point-Mei determining system for Eq.~\eqref{eq:L}; then, if additional derivative-dependent candidates are of interest, formulate and solve the generalized determining equations at a specified differential order.  This separation prevents point and generalized symmetries from being counted in a single algebra without a common prolongation framework.

\subsection{Physical interpretation of the scalar-flat branch}\label{sec:scalarflat}

The degeneracy of the $R=0$ sector is a distinctive feature of pure quadratic gravity.  Substitution of a constant $R=0$ into the vacuum equations gives $f=f_R=0$ and removes the Einstein-like algebraic term that would otherwise constrain $R_{\mu\nu}$.  Consequently, scalar flatness does not imply Ricci flatness in this branch.  This is why a metric with the Reissner--Nordstr\"om functional form can occur without an electromagnetic stress tensor.

It is important here to comment on the physical interpretation.  In Einstein--Maxwell theory the coefficient of $r^{-2}$ is tied to the Maxwell charge through the matter field equations and normalization of the electromagnetic action.  In the pure $R^2$ model considered here, the integration constant
$C$ appearing in the Reissner--Nordstr\"om-form scalar-flat solution
has no such identification.  Calling $C$ an electric charge would therefore import information that is absent from the action.  The safest terminology is ``Reissner--Nordstr\"om-form scalar-flat solution'' unless a Maxwell sector is added explicitly.

This observation also separates two different notions of black-hole reconstruction.  One may reproduce the functional form of a familiar metric within a modified gravitational vacuum, or one may reproduce the full coupled gravitational and matter solution with the same physical interpretation of its integration constants.  Only the second warrants identifying $C$ with an electromagnetic charge.

\section{Scope, reproducibility, and open problems}\label{sec:scope}

The results above are deliberately limited to statements that can be checked directly: the corrected variational equations, the stated Noether residuals, the curvature flux, the displayed strong point-Mei tests, and the second-prolongation Lie classification in the polynomial point class of total degree at most two. The accompanying symbolic audit reproduces each result from Eq.~\eqref{eq:L}.

Three extensions would materially strengthen the classification.  First, the full point-Noether determining PDEs should be solved without an affine ansatz, including a general gauge function $K(r,H,N,R)$.  Second, the unrestricted point-Mei determining system should be solved for the corrected Lagrangian.  Third, generalized derivative-dependent Mei candidates should be tested using a genuine generalized prolongation rather than the point formulas.  These are separate mathematical problems and should not be conflated with verification of the generators displayed in the present work.

On the solution side, the main open problem is the global $C_R\neq0$ sector.  The covariant first integral \eqref{eq:unfixedflux} reduces its order and supplies a useful shooting parameter, but regular horizons, asymptotic behavior, and global causal structure require the remaining field equations.  A numerical study in the unfixed areal-radius variables $(A,B,R)$ would therefore be a natural continuation.  The asymptotic relation \eqref{eq:Rasymptotic} can be used as one consistency check on such integrations.

\section{Conclusions}\label{sec:conclusion}

We have examined Noether, Mei, and Lie symmetries of static spherical pure $R^2$ gravity from a radial action derived before the gravitational constraint is removed. The decisive term is $2f_R=4R$. Although it is invisible to the velocity Hessian, it contributes to the variation. Retaining it yields Eq.~\eqref{eq:L}, whose Euler--Lagrange equations reproduce the geometric Ricci scalar.

Radial translation and the combined scaling in Eq.~\eqref{eq:XN} are exact off-shell Noether symmetries. The latter gives $I_N=6HNR'-rE_L$. Restoring the metric equation removed by the radial gauge supplies $E_L=0$, and the charge reduces to $HNR'=C_R$.

The four-dimensional trace equation supplies an independent derivation: in pure $R^2$ gravity, $\Box R=0$, whose static spherical first integral is the same flux. This agreement both checks the radial reduction and identifies the covariant origin of the Noether charge.

We have also shown that Noether, strong Mei, and Lie symmetries have to be treated as distinct notions.  In particular, the generator $\tfrac12r\partial_r+H\partial_H$ satisfies the strong point-Mei condition but does not satisfy the ordinary Noether condition, whereas the combined scaling \eqref{eq:XN} is an exact Noether symmetry.  Both transformations belong to the dynamical Lie algebra.  By considering the complete polynomial point ansatz of total degree not exceeding two, containing 60 independent coefficients, we obtain a determining matrix of rank 57.  Therefore the polynomial Lie-point algebra in this class is three-dimensional and is isomorphic to $\mathfrak{aff}(1)\oplus\mathbb R$.  This result excludes additional constant, linear, and quadratic polynomial point generators, although it does not exclude genuinely non-polynomial or generalized symmetries.

The conserved curvature flux also gives a useful classification of the solution space.  For $C_R=0$, the scalar curvature is constant in every regular region.  The nonzero constant-curvature branch reduces to an Einstein space and gives the Schwarzschild--(anti-)de Sitter family.  The $R=0$ branch is instead degenerate in pure quadratic gravity because both $f(0)$ and $f_R(0)$ vanish.  Consequently, scalar flatness does not imply Ricci flatness, and the Reissner--Nordstr\"om metric form can occur without a Maxwell energy-momentum tensor.  In this case the coefficient of $r^{-2}$ is only an integration constant and cannot be identified with an electric charge unless an electromagnetic sector is introduced explicitly.

For $C_R\neq0$, the scalar curvature is dynamical and satisfies $R'=C_R/(HN)$ in the gauge-fixed variables, or the equivalent unfixed relation \eqref{eq:unfixedflux}.  This branch has a qualitatively different structure from the constant-curvature sector.  The curvature is locally monotonic on regular connected regions, and its asymptotic behavior is controlled by the same flux constant.  A complete study of regular horizons and global causal structure requires solving the remaining metric equations and is left for future work.

The Lie result is complete only within the polynomial point class of total degree at most two; the unrestricted non-polynomial problem remains open. Derivative-dependent Mei candidates likewise require a separate generalized prolongation. Together with a numerical treatment of the nonzero-flux sector, these questions are left for future work.

\section*{Data and code availability}

No numerical or observational dataset is used in this work.  The symbolic
results are accompanied by the SymPy script
\texttt{R2\_final\_reproducibility\_audit.py}.  The script independently
derives the Euler--Lagrange normalization, canonical momenta, Hessian
determinant, radial energy, Noether residuals, first integral, strong
point-Mei residuals, curvature-flux identity, Lie brackets, and the
total-degree-$\leq2$ polynomial Lie determining system.  A clean execution
returns 33 successful core checks with zero failures and reconstructs the
$495\times60$ determining matrix with rank 57 and a three-dimensional
nullspace.  The code is intended to be supplied as supplementary material
with the submitted manuscript.

\section*{Acknowledgments}
This work was supported and funded by the Deanship of Scientific Research at Imam Mohammad Ibn Saud Islamic University (IMSIU) [grant number IMSIU-DDRSP2602]. The work of Kazuharu Bamba was supported in part by the JSPS KAKENHI Grants No.~24KF0100, No.~25KF0176 and a grant-in-aid of academic research of the Yamaguchi Scholarship Foundation. 

\appendix
\section{Symbolic verification identities}\label{app:verification}
This appendix summarizes the symbolic identities used to verify the principal results of the manuscript. It provides reproducible checks of the Euler--Lagrange equations, canonical quantities, Noether and Mei symmetry conditions, curvature-flux relation, and polynomial Lie-point classification. Starting from Eq.~\eqref{eq:L}, direct differentiation gives
\begin{align}
 \mathcal E_H(L)&=\frac{1}{N}\left[2N^2R''+2NN''R-R(N')^2\right],\quad
 \mathcal E_N(L)=\frac{1}{N^2}\left[\text{left-hand side of Eq.~\eqref{eq:EN}}\right],\nonumber\\
 \mathcal E_R(L)&=\frac{1}{N}\left[\text{left-hand side of Eq.~\eqref{eq:ER}}\right].
\end{align}
The remaining identities used as automatic zero-residual tests are
\begin{align}
 \det W-24HNR&=0,\quad
 X_N^{[1]}L+L=0,\quad
 Y_2^{[1]}L+L-(4R-NR^2)=0,\quad
 Y_3^{[1]}L-E_L=0,\nonumber\\
 Hp_H+Np_N-Rp_R-6HNR'&=0,\quad 
 I_N-(6HNR'-rE_L)=0,\quad 
 M_2^{[1]}L=0,\quad 
 \frac{\dd}{\dd r}(HNR')-N\Box R=0.
\end{align}
For the Lie tests, the second prolongation is applied to Eqs.~\eqref{eq:EH}--\eqref{eq:ER}  on the regular sector \(HNR\neq0\), where the system can be solved for \((H'',N'',R'')\), and the resulting expressions are then restricted to the solution manifold.

For the quadratic polynomial classification of Sec.~\ref{sec:lie}, the reproducibility calculation uses the complete monomial basis \eqref{eq:quadbasis} independently in all four generator components.  The symbolic workflow produces 893 raw coefficient conditions, 495 distinct linear equations, and a $495\times60$ coefficient matrix of rank 57.  Its three-dimensional nullspace is spanned by
$\{L_1,L_2,L_4\}$, equivalently by the basis
$\{L_1,L_2,L_3\}$ introduced in the main text. Substitution of each nullspace generator back into the full second-prolongation equations gives identically vanishing on-shell residuals.  These rank and nullspace checks establish completeness within the degree-$\leq2$ polynomial point class; they do not exclude non-polynomial point generators.

The accompanying SymPy audit reports zero residuals for the implemented identities asserted in the main text and independently reconstructs the polynomial Lie determining matrix.  In a clean run it gives 33 successful core checks and no failures, followed by the independent classification checks
\[
60\ \text{unknowns},\qquad
893\ \text{raw conditions},\qquad
495\ \text{distinct conditions},\qquad
\operatorname{rank}M=57,\qquad
\dim\ker M=3.
\]
The nullspace is therefore three-dimensional and may be represented
by either basis $\{L_1,L_2,L_4\}$ or
$\{L_1,L_2,L_3\}$.

\section{Variational status of the potential term}\label{app:potential}

This appendix explains why a term that is independent of the radial velocities cannot be discarded solely because it does not affect the Hessian. Although such a term leaves the reduced Legendre map unchanged, it contributes to the Euler--Lagrange equations and can therefore alter the symmetry conditions and conserved quantities. To demonstrate this point, consider
\begin{equation}
 \widetilde L=L_0+2f_R(R).
\end{equation}
The Hessians of $\widetilde L$ and $L_0$ with respect to $(H',N',R')$ are identical, but their $R$ Euler--Lagrange expressions differ.  With the convention used in Sec.~\ref{sec:eom},
\begin{equation}
 \mathcal E_R(L)=D_r\!\left(\frac{\partial L}{\partial R'}\right)-\frac{\partial L}{\partial R},
\end{equation}
one has
\begin{equation}
 \mathcal E_R(\widetilde L)-\mathcal E_R(L_0)=-2f_{RR}.
\end{equation}
With the opposite Euler--Lagrange sign convention the difference is $+2f_{RR}$.  Thus Hessian equivalence does not imply variational equivalence.  Only a genuine total derivative can be removed without changing the bulk Euler--Lagrange equations.

\end{document}